\documentclass[twocolumn,aps,prl,amsmath,showpacs]{revtex4}
\usepackage{graphicx}
\usepackage{dcolumn}%
\usepackage{bm}%
\usepackage{amssymb}
\usepackage{amsmath}
\usepackage{color}

\def\be{\begin{equation}}
\def\ee{\end{equation}}

\begin{document}

\title{Elasticity of arrested short-ranged attractive colloids:\\
homogeneous and heterogeneous glasses}
\author{Alessio Zaccone$^{1}$, Hua Wu$^{1}$
and Emanuela Del Gado$^{2}$} \affiliation{${}^1$Chemistry and
Applied Biosciences, ETH Z\"urich, CH-8093 Z\"urich, Switzerland}
\affiliation{${}^2$Polymer Physics, ETH Z\"urich, CH-8093 Z\"urich,
Switzerland}
\date{\today}
\begin{abstract}
We evaluate the elasticity of arrested short-ranged attractive
colloids by combining an analytically solvable elastic model with a
hierarchical arrest scheme into a new approach, which 
allows to discriminate the microscopic (primary
particle-level) from the mesoscopic (cluster-level) contribution to
the macroscopic shear modulus. The results quantitatively predict
experimental data in a wide range of volume fractions and indicate
in which cases the relevant contribution is due to mesoscopic
structures. On this basis we propose that different arrested states
of short-ranged attractive colloids can be meaningfully
distinguished as homogeneous or heterogeneous colloidal glasses in
terms of the length-scale which controls their elastic behavior.
\end{abstract}
\pacs{61.43.-j,82.70.Dd,81.40.Jj}

\maketitle Solutions of short-ranged attractive colloidal particles
are the object of intense study due to their technological applications
(proteins, paints etc.), as the constituent blocks of
nanomaterials, as well as model systems to
investigate
phase behavior and dynamical arrest of condensed
matter~\cite{frenkel}. A landscape of phases has been observed upon
varying the volume fraction $\phi$ or the interaction
parameters~\cite{poon}, and, as a matter of fact, extended regions
of the phase diagram are still poorly understood. In very dense
suspensions ($\phi > 0.5$) the arrested states are spatially
homogeneous (i.e. the typical linear size of structural
heterogeneity is smaller than the particle diameter $R_{0}$).
Particles are immobilized within the range of attraction,
giving rise to bonds that are persistent under strain
in the linear regime, and the high density leads to
(attractive) glassy states~\cite{attr_glasses}. For $0.2 < \phi < 0.5$, 
the situation is complicated:
arrested states can only occur thanks to pronounced
structural heterogeneities, typically on length scales larger than
$R_{0}$. Therefore, they are more
related to gelation~\cite{ema2}, 
rather than to the caging typical of crowded
random media~\cite{fuchs08}. 
However, such arrested states are also hardly 
classifiable as classic network gels
in view of the different morphology and stress-bearing mechanisms.
For them, different microscopic phenomena should be considered and
a new theoretical framework, able to account for the strong spatial
heterogeneities and currently still missing, would be desirable.
A crucial point, mostly neglected in recent studies,
is that arrested states occurring at different volume fractions
and attraction strengths do display dramatically
diverse mechanical and rheological properties~\cite{exp}.
That is why, their characterization is of true interest in
technological applications and material design.

In this Letter we propose a new, more 
{\it down to earth}, approach to
characterize arrested short-ranged attractive colloids, based on
their mechanical response. 
Being the problem extremely hard to tackle, 
we rely on a simplified picture: We combine an analytically solvable
elastic model with a hierarchical arrest scheme and 
provide a first attempt to discriminate the
microscopic (primary particle-level) from the mesoscopic
(cluster-level) contribution to the macroscopic elasticity. The
predictions are given in terms of the shear modulus, a quantity that
can be measured in a rheometer, and are directly compared with
experimental results. In spite of the crude, but physically grounded 
assumptions, we are able to show, for the first time, that
where structural heterogeneities as clusters are present, the measured 
macroscopic elasticity differs by orders of magnitude from the
one of the homogeneous glass because it is strongly dominated by 
the inter-cluster contribution. 
Our approach offers therefore a new insight into the Physics of arrested 
attractive colloidal suspensions, especially in the
range $0.2<\phi<0.5$, hitherto poorly understood. 
We propose that homogeneous and heterogeneous arrested
states can be distinguished in terms of the length scale
$\tilde{R}_{0}$, which controls the macroscopic elastic modulus.

{\it Model and predictions.} We study a suspension
of colloidal particles interacting via a short-range attraction, as
the one typically induced by depletion using non adsorbing polymer
\cite{frenkel}, well above percolation, so that a
finite shear modulus is always detectable.
As a first step towards rationalizing the mechanical response of 
arrested, dense attractive colloids, and in the fundamental lack,
at present, of alternative approaches, we use a simplified, although 
physically grounded, scenario.
We consider that, by reducing the volume fraction from the very dense 
homogeneous glass case, the attractive interactions will start to 
produce small aggregates, with the aggregation being soon arrested by the 
glassy environment. This mechanism leads to compact aggregates of 
relatively small linear size~\cite{zukoski,laurati}. 
In order to distinguish
the single-particle contribution to elasticity from the mesoscopic one, 
we therefore consider coarse-grained entities, 
i.e. {\em renormalized particles}. 
Their effective interactions are directly responsible for the macroscopic
properties of the system. This scheme can
be associated to a double ergodicity-breaking
scenario where the arrest occurs in form of a
cluster-glass transition~\cite{segre,kroy,ema1}.
A {\it sine qua non} is that further coalescence of the clusters,
leading to phase separation, is prevented on the observation
time scale~\cite{kroy}.
We distinguish between the macroscopic elasticity of the system
$G$, the intra-cluster elasticity $G_{g}$ resulting from
mutual interactions between primary particles, and the inter-cluster
elasticity $G_{c}$ resulting from mutual interactions between
clusters.

Let us first consider the case of a homogeneous attractive glass in
the low-temperature and high-density ($\phi > 0.5$) region of the
phase diagram. Here we focus on the first linear regime reported
in rheological measurements~\cite{pham}, which is due to bond-breaking
and, for strong attractions, extends nearly up to strains of a few percent.
Hence, we follow the Cauchy-Born approach developed in~\cite{alexander}
for amorphous solids and obtain the elasticity
from a free energy
expansion around a \emph{stressed} (quenched) reference state $\{R\}$ where
all particles are {\it labeled}.
The expansion reads $\delta F \equiv F(\{r\}) - F
(\{R\}) \simeq \sum_{ij} \left[
\partial F /
\partial r_{ij} \right] \delta r_{ij} + \frac{1}{2} \sum_{ij,kl}
\left[ \partial^{2} F/ \partial r_{ij} \partial r_{kl} \right]
\delta r_{ij}\delta r_{kl}$, where the sums run over all interacting
pairs of particles (bonds) and the derivatives are evaluated at
relative distances $\{\mathbf{R}_{ij}\}$ in the quenched reference
state. Due to the purely internal nature of the stresses, the first
term in the r.h.s. does not contribute to the macroscopic
elasticity~\cite{alexander,barrat}. In the case of central forces,
considering a pair-interaction with a deep minimum $\epsilon \gg
k_{B}T$, the bond stiffness is defined by $\kappa\simeq\partial^{2}
F/\partial r_{ij}^{2}$, and the Born-Huang (BH) terms become $\delta
F = \frac{1}{2}\kappa \sum_{ij} (\delta r_{ij})^{2}$.
If we express such deviations in terms of a smooth displacement
field $\mathbf{u}_{ij}$ \cite{alexander}, with central forces only
the longitudinal component $u_{ij}^{\parallel}$ is non-zero so that,
upon neglecting higher order terms, the deformation free energy
reads $\delta F \simeq \frac{1}{2}\kappa \sum_{ij}
(u_{ij}^{\parallel})^{2}$. Under the assumption that
all pairs of particles are much localized near the minimum of
the potential well and that displacements are affine, 
we obtain the continuum limit of the microscopic
$u_{ij}^{\parallel}$ in terms of the macroscopic
strain tensor $\mathbf{e}$, $\langle
u_{ij}^{\parallel} \rangle= Tr[(\mathbf{R}_{ij} \otimes
\mathbf{R}_{ij}) \cdot \mathbf{e} /R_{ij} ]$, with $\langle
u_{ij}^{\parallel}-\langle u_{ij}^{\parallel}\rangle\rangle^{2}\ll
\langle u_{ij}^{\parallel}\rangle^{2}$, to obtain $\delta F \simeq
\frac{1}{2}\kappa \sum_{ij} \{ Tr[(\mathbf{R}_{ij} \otimes
\mathbf{R}_{ij}) \cdot \mathbf{e} /R_{ij} ]\}^{2}$. For the quenched
reference configuration $\{R\}$, in the case of uncorrelated
disorder, the summation over bonds can be replaced by the total
number of contacts, which introduces the average coordination number
$z(\phi)$ in the final expression of the deformation free energy. We
then derive the off-diagonal components of the stress
tensor~\cite{alessio0}
\begin{equation}
\sigma_{\alpha \beta} \equiv \frac{\partial \delta F} {\partial
e_{\alpha \beta}} \simeq 2\rho z(\phi) \kappa
\frac{\partial}{\partial e_{\alpha \beta}} \left\langle
\left(\frac{R^{\alpha}_{ij}R^{\beta}_{ij}}{R_{ij}}\right)^{2}e_{\alpha
\beta}^{2} \right\rangle_{\Omega}
\end{equation}
where $\left\langle\bullet\right\rangle_{\Omega}=(4 \pi)^{-1}
\int\bullet d\Omega$. Hence the shear modulus of an arrested phase
of volume fraction $\phi= \rho\pi R_{0}^{3}/6$, with mean particle
diameter $\simeq R_{0}$, is given by $G_{g} = (2/5) \pi^{-1} \phi
z(\phi) \kappa R_{0}^{2-d}$. The evaluation of the 
coordination number $z(\phi)$ in dense glassy systems is still controversial.
Since at high densities also 
attractive systems are likely to be dominated by excluded-volume 
repulsion~\cite{kaufman}, we estimate the number of short-range contacts 
involved in the mechanical bonds from $z(\phi)$ of a hard sphere glass. 
That is, we integrate the radial distribution function (rdf) of the dense 
(hard-sphere) liquid precursor, $g(r)$, within a shell of width 
$l^{\dagger}$. A reasonable choice for repulsive systems is 
$l^{\dagger}\sim0.03$ which allows to recover $z\sim6$ at random 
close packing~\cite{alessio}.  
We thus obtain the shear modulus as a function of the rdf:
\begin{equation}
G_{g} \simeq (48/5) \pi^{-1} \kappa R_{0}^{2-d} \phi
\int_{0}^{l^{\dagger}} (1+l)^{2} g(l;\phi) dl \label{eqmod2}
\end{equation}
where $l=(r-R_{0})/R_{0}$ and $l^{\dagger}\simeq 1/30$.
For $g(r)$ near contact ($l<0.1$) we use
liquid theory
valid in the dense hard-sphere fluid~\cite{Henderson} and we
calculate $\kappa$ using the Asakura-Oosawa (AO) potential~\cite{AO}.
Our predictions have been compared to the experimental data of
Ref.~\cite{pham} for an attractive colloidal glass ($\phi \simeq
0.6$). The shear modulus was measured 
as a function of the attraction strength, i.e. the reduced polymer
concentration $c_{p}/c_{p}^{*}$. 
For strong enough attraction (and fully elastic response), the
affine approximation is more realistic and our prediction gives an 
accurate estimate of the measured shear
modulus~\cite{suppl}. We have checked that the agreement does not 
significantly change upon vary $l^{\dagger}$.

For a more stringent test of the model
where the bond-stiffness and the local structure
of the glass are well-defined, we
performed Molecular Dynamics simulations of deeply quenched glasses
obtained from different models for supercooled liquids. 
We subjected the glass to shear strain and extracted stress-strain
curves~\cite{suppl}. The agreement obtained 
in the linear regime is remarkable, indicating that the model is 
quantitatively predictive and 
that the nonaffine rearrangements are actually small in the 
strong-attraction or deep-quench limit in agreement with our assumptions
~\cite{suppl}.

For the more dilute regime, 
we now consider local aggregation of the colloidal particles to form beads
changed by small variation of $l^{\dagger}$.
(clusters) which, in turn, arrest due to either caging or residual
attraction, as in the double-ergodicity breaking
scenario mentioned above~\cite{kroy,ema1,segre}.
The clusters are viewed as compact (spherical or quasi-spherical)
renormalized particles of diameter $\tilde{R_{0}}$, 
whose effective volume fraction may be identified with the one
determined by the spheres enclosing them (i.e. significantly
larger than $\phi$). If the cluster linear size is larger than the
particle diameter by a factor say less than 10, each contact between
clusters is likely to reduce to a single colloid-colloid bond. Upon
neglecting: 1) the breakup probability within the cluster, and 2)
the effect of long-range repulsion, the effective interaction
between clusters obviously reduces to the bare colloid-colloid
interaction, in agreement with~\cite{kroy}. Further, the mean
coordination will change to $z(\phi_{c})$ (where now $\phi_{c}$ is
the cluster volume fraction), but its form can be still determined
as in Eq.(\ref{eqmod2}) if $\phi_{c}$ is in the dense glassy
regime dominated by mutual impenetrability. 
Hence, for the modulus of the cluster-glass we can write:
\begin{equation}
G_{c} \simeq (2/5) \pi^{-1} \phi_{c} z (\phi_{c}) \tilde{\kappa}
\tilde{R_{0}}^{2-d}\label{eqmod3}
\end{equation}
with $\tilde{\kappa} \simeq \kappa$ under the assumption of small
clusters.
Eq. (\ref{eqmod3}) gives the elastic modulus of the material, where the
macroscopic elasticity is dominated by the mesoscopic level.
\begin{figure}
\includegraphics[width=1.0\linewidth]{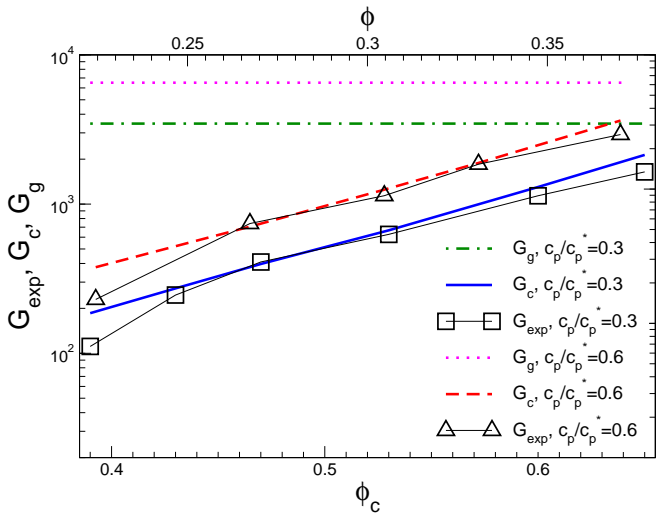}
\caption{(color online) Shear modulus as a function of $\phi_{c}$
for the experimental data from~\cite{zukoski} and for the
model prediction from Eqs.(\ref{eqmod2}) and (\ref{eqmod3}). 
$\tilde{R_{0}}/R_{0}=5.5$ and polymer-to-colloid size ratio $\xi=0.078$
from~\cite{zukoski} (upper x-axis: $\phi$ for $c_{p}/c_{p}^{*}=0.3$).
 } \label{fig2}
\end{figure}
We test our scheme using the extensive experimental data of
Ref.~\cite{zukoski} for a system of colloidal silica particles with
polystyrene as depletant in organic solvent (decalin), 
where the ratio of the polymer
gyration radius to $R_{0}$ is the same as in the experiments
of~\cite{pham}. For two values of $c_{p}/c_{p}^{*}$, they vary $\phi$ 
in the range $0.2-0.4$. The system is a dense gel of compact clusters 
($\phi \simeq 0.5$ inside the cluster), whose diameter 
$\tilde{R_{0}}/R_{0} \sim 5$ is determined from
the Debye-Bueche plot. For a fixed 
$c_{p}/c_{p}^{*}$, they find $\tilde{R_{0}}/R_{0}$ and the volume fraction 
inside the clusters not to significantly change with $\phi$, i.e. $\phi_{c}$ 
increasing linearly with $\phi$. 
This indicates that in this case our assumptions on the structure are 
reasonable.
The assumption that the spheres enclosing 
the clusters are densely 
packed yields $\phi_{c}\simeq 0.64$ at the largest value of $\phi$, 
allowing for a physically reasonable estimate of $\phi_{c}$ in the range 
of $\phi$ considered. 
Using the experimental measurement of $\tilde{R_{0}}/R_{0}$ 
we can therefore calculate $G_{c}$ from Eq.(\ref{eqmod3}). 
In Fig.\ref{fig2} we plot the values of the shear modulus measured at 
1Hz in~\cite{zukoski} for two different attraction strengths, together 
with our predictions of $G_{g}$ from Eq.(\ref{eqmod2}) \cite{footnote}
and of $G_{c}$ from Eq.(\ref{eqmod3}), as function of $\phi$ and 
$\phi_{c}$. 
The excellent agreement between our estimate and experiments in 
Fig.\ref{fig2} clearly indicates that, differently
from the case previously considered, now is $G_{c}$ that dominates
the macroscopic elasticity of the system. 
We have checked that the uncertainty on $\phi_{c}$ does not significantly 
affect this result.
For lower attractions, the system will gradually cross over
towards the repulsive case.
\begin{figure}
\includegraphics[width=1.0\linewidth]{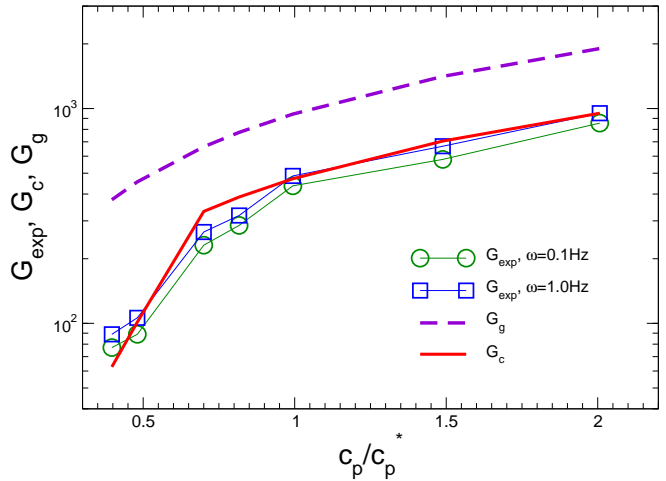}
\caption{(color online) The shear modulus
as a function of $c_{p}/c_{p}^{*}$ from Ref.\cite{laurati} (symbols), 
with polymer-to-colloid
size ratio $\xi=0.08$ and $\tilde{R_{0}}/R_{0}$ from experiments.
Lines: model prediction from Eq.(\ref{eqmod2}) (dashed) and 
(\ref{eqmod3}) (full).
}
\label{fig3}
\end{figure}

We now consider the results recently reported in \cite{laurati},
for a different arrested AO system
in the heterogeneous glass regime, studied at a fixed $\phi=0.4$ and upon 
varying $c_{p}/c_{p}^{*} \simeq 2$. 
In the experiments, they used light-scattering and microscopy to
estimate the length-scale of structural heterogeneities up to 
$c_{p}/c_{p}^{*} \simeq 2$.
We have used this experimental input to calculate the prediction 
of our model for this system. Again for densely packed clusters 
(i.e. with volume fraction in the range 
$0.58$-$0.64$) we estimate $\phi_{c}\sim 0.6$ (and check that variation 
around this value does not significantly change our outcomes). 
In Fig.\ref{fig3} we compare $G_{c}$ to the experimental measurements:
$G_{c}$ (solid line) shows a remarkable agreement 
with $G_{exp}$ and turns out again to dominate the elastic response 
(see in comparison $G_{g}$, dashed line).
Also here the comparison with experiments has been done at
low frequencies and for fully elastic response.

\begin{figure}
\includegraphics[width=0.8\linewidth]{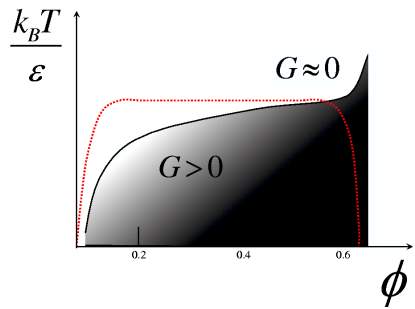}
\caption{(color online) Arrested states characterized in terms of 
the typical length scale $\tilde{R}_{0}/R_{0}$ of the mechanically 
relevant structural heterogeneities. 
Color gradient from $\tilde{R_{0}}/R_{0}\gg1$ (light)
to $\tilde{R_{0}}/R_{0}\simeq1$ (dark). 
Fluid-solid transition (solid line) and 
spinodal (dotted line) also indicated.}
\label{fig4}
\end{figure}

The results just discussed lead us to sketch in Fig.\ref{fig4} a new
qualitative {\it phase diagram} for the arrested states of
short-range attractive colloidal suspensions. As a function of
$\phi$ and $k_{B}T/\epsilon$, where $\epsilon$ is the depth of the
attractive well, we locate the arrested states in the region where
the system displays a non-zero measurable elastic modulus $G$ (the
continuum line divides fluid from solid states as discussed e.g.
in~\cite{trappe}). At high enough $\phi$ and large attractions, such
{\it state} ({\it homogeneous glass}) is characterized by an elastic
modulus dominated by the inter-particle elasticity
($\tilde{R_{0}}/R_{0}=1$). Upon lowering $\phi$, aggregation starts
to produce mesoscopic structural heterogeneities and the mechanical
response of the system crosses over towards a regime dominated by
the inter-cluster elasticity ({\it heterogeneous glass})
($\tilde{R_{0}}/R_{0}\gg1$). The two regimes will be distinguished
by a significant variation of the elastic modulus of the material
(Fig.\ref{fig2}).
Our diagram suggests for the first time a
distinction between arrested states of attractive colloidal
suspensions in terms of a well defined, directly measurable
quantity, the elastic shear modulus.
When the structure of the material is known in some detail,
such distinction can provide information on which part of the
structure is relevant to the elastic behavior. 
Upon further lowering the volume
fraction ($\phi\lesssim 0.2$), we expect the elastic response to be
dominated by a different length scale, associated to the weakly
connected network-like mesoscopic structure and strongly dependent
on $\phi$ \cite{ema2}. Due to lower connectivity, nonaffinity is
likely to be more pronounced and lead to a sensibly lower shear
modulus (Fig.\ref{fig2}). At very low $\phi$ arrested states may be
due to effective directional interactions arising at mesoscopic
length scales, leading to open network structures~\cite{network},
where buckling strongly affects the macroscopic elasticity.

{\it Conclusions.} 
We have proposed that dramatic differences observed at different volume 
fractions in the mechanical response of arrested, dense attractive colloids
originate from the presence of structural heterogeneities. 
We have presented a first attempt to estimate the shear modulus of 
glassy systems where structural heterogeneities as clusters are present. 
Being the problem extremely hard to tackle, we have to rely on a rather 
simplified picture. In spite of this, we have been able to show 
in two different experimental systems that the measured 
macroscopic elasticity is strongly dominated by the 
inter-cluster contribution. 
This first novel approach can pave the way to a
meaningful distinction between different glassy states, on the basis
of the length-scale $\tilde{R}_{0}$ dominating their elastic
response. $\tilde{R}_{0}$ varies with volume fraction and attraction strength 
due to the
arising of structural heterogeneities. However, whereas the
characterization of structural heterogeneities requires detailed
structural information and is therefore often elusive, the variation
of $\tilde{R}_{0}$ is unambiguously signalled by significant
variation of the elastic modulus. 
Therefore we propose that homogeneous and heterogeneous arrested
states can be distinguished in terms of the length scale
$\tilde{R}_{0}$, which controls the macroscopic elastic modulus.


\begin{thebibliography}{99}
\bibitem{footnote}
$G_{g}$ in Fig.\ref{fig2} has been calculated from Eq.(\ref{eqmod2})
taking the volume fraction inside the clusters $\simeq 0.5$ and
independent on $\phi$~\cite{zukoski}.
\end{thebibliography}
\end{document}